\documentclass[preprint,showpacs,preprintnumbers,amsmath,amssymb]{revtex4}

\usepackage{graphicx}
\usepackage{dcolumn}
\usepackage{bm}

\newcommand{\etal}{\mbox{\em et al.}}

\begin{document}


\title{Proton elastic scattering from tin isotopes at 295~MeV and 
systematic change of neutron density distributions}
\author{
S.~Terashima$^1$\footnote[1]
{Present address: RIKEN Nishina Center, Wako, Saitama 351-0198, Japan },
H.~Sakaguchi$^1$\footnote[2]
{Present address: Miyazaki University, Miyazaki, Miyazaki 889-2192, Japan },
H.~Takeda$^{1\ast}$, 
T.~Ishikawa$^1$\footnote[3]{Present address: Laboratory of Nuclear Science, 
Tohoku University, Sendai, Miyagi 982-0216, Japan }, 
M.~Itoh$^1$\footnote[4]{Present address: Cyclotron and Radioisotope Center, 
Tohoku University, Sendai, Miyagi 980-8578, Japan }, 
T.~Kawabata$^1$\footnote[5]{Present address : Center for Nuclear Study, 
University of Tokyo, Wako, Saitama 351-0198, Japan },
T.~Murakami$^1$,
M.~Uchida$^1$\footnote[6]{Present address : Tokyo Institute of Technology, 
Ookayama, Tokyo 152-8550, Japan },
Y.~Yasuda$^1$\footnote[7]{Present address : University of Tsukuba, 
Tsukuba, Ibaraki 305-8571, Japan },
M.~Yosoi$^{1,2}$, 
J.~Zenihiro$^1$, 
H.~P.~Yoshida$^{2\S}$, 
T.~Noro$^3$,
T.~Ishida$^{3\ddag}$,
S.~Asaji$^3$,
T.~Yonemura$^3$
}
\address{
$^1$ \it Department of Physics, Kyoto University, 
Kyoto 606-8502, Japan \\
$^2$ \it Research Center for Nuclear Physics (RCNP), Osaka University,
Osaka 567-0047, Japan \\
$^3$ \it Department of Physics, Kyushu University, Fukuoka 812-8581 Japan 
}

\date{\today}

\begin{abstract}                
Cross sections and analyzing powers for proton elastic 
scattering from $^{116,118,120,122,124}$Sn  at 295 MeV have been 
measured for a momentum transfer of up to about 3.5 fm$^{-1}$ 
to deduce systematic changes of the neutron density distribution.
We tuned the relativistic Love-Franey interaction to explain the proton 
elastic scattering of a nucleus whose density distribution is well known. 
Then, we applied this interaction to deduce the neutron density 
distributions of tin isotopes. 
The result of our analysis shows the clear systematic behavior of a gradual 
increase in the neutron skin thickness of tin isotopes with mass number.
\end{abstract}

\maketitle
\section{Introduction}               
Charge distributions in stable nuclei have been reliably measured 
by electron elastic scattering and muonic x-ray data~\cite{deVries}.
These charge-sensitive experiments have provided precise 
information on charge distributions.
On the other hand, it 
is much more difficult to deduce neutron density distributions, 
since electromagnetic interaction provides little
information on neutron density distributions. 
The proton and neutron density distributions have a similar shape in stable 
nuclei. However in recent research it has been reported that in some 
unstable nuclei the differences between proton and neutron shapes are 
greater than those in stable nuclei~\cite{TAN}.  
It has also been indicated that the thickness of the neutron skin is 
closely related to the symmetry term of 
the equation of state~(EOS)~\cite{TYP,DAN}.
Thus, the determination of neutron density distributions has become 
increasingly important. 

There have been many experiments attempting to extract 
neutron and matter density distributions 
in the nuclear interior using hadronic probes~\cite{Batty}.
Pion and alpha elastic scattering have been performed 
in the study of neutron and matter density distributions~\cite{GIL,JOH,BAR}.
Compared with other hadronic probes, 
the elastic scattering of protons at intermediate energies is suitable 
for extracting information on the nuclear surface and interior, 
because at intermediate energies, proton elastic scattering has a simple 
reaction mechanism. 
To deduce nuclear densities using protons, 
the incident energy has to be sufficiently high to describe the scattering 
by the simple mechanism. 
At energies above 100~MeV, we can explain proton elastic scattering 
microscopically, because the imaginary part of the optical potential 
describes a quasi-free process mainly without the need for a 
renormalization factor. So far, energies above 500~MeV have been 
applied for proton elastic scattering to study neutron density 
distributions~\cite{Ray,Sta}. However, this energy is sufficiently 
high to produce mesons, and information on the nuclear interior is 
easily masked by the meson-productions. 
Furthermore, the total cross section of nucleon-nucleon 
scattering shows a minimum at the incident energy of 300~MeV.
We thus adopt 300 MeV protons in this work as probes for information on 
the nuclear interior. 

In our previous papers~\cite{SAK}, we tuned the effective relativistic 
Love-Franey interaction by introducing 'so-called medium effects' 
for the scattering from a nucleus whose density distribution is 
well known. We used elastic scattering from $^{58}$Ni
to tune the interaction, since $^{58}$Ni is the heaviest stable nucleus 
with N $\approx $ Z and the density distribution of the neutrons in
$^{58}$Ni can be assumed to be the same as that for the protons.
To explain the results of our experiments we found that we had 
to modify the scattering amplitudes of the nucleon-nucleon interactions 
inside the nucleus as follows. We phenomenologically changed the masses
and coupling constants of exchanged mesons depending on the
nuclear density. 
Thus, we could explain the scattering sufficiently well to deduce 
the matter density distribution precisely.

For our first systematic search for neutron density distributions, 
we selected tin isotopes. Tin has many stable isotopes
~($^{112}$Sn-$^{124}$Sn). Also, unstable tin isotopes have a long 
isotopic chain including two double-magic nuclei
~($^{100}$Sn~[N=50], $^{132}$Sn~[N=82]). Moreover, its proton number is 
a magic number~(Z=50). Thus, tin isotopes are suitable for the study of 
systematic changes in neutron density distributions. The main purposes 
of this work are to attempt to deduce information on neutron density 
distributions, and to systematically study the neutron skin thickness 
of tin isotopes. The experimental setup and procedure are described 
in Sec. II. Details of the analysis used to deduce neutron density 
distributions are given in Sec. III. The deduced radii of tin isotopes 
are discussed in Sec. IV.A summary is given in Sec. V.

\section{Experiment}
The measurements were performed at the Research Center for Nuclear
Physics (RCNP), Osaka University. 
Polarized protons from a high-intensity polarized ion source 
were injected into an AVF cyclotron, transported to
a six-sector ring cyclotron and accelerated up to 295 MeV. The
polarization axis was in the vertical direction. The spin direction and
magnitude of the beam polarization were measured continuously using 
sampling-type beam-line polarimeters (BLPs)~\cite{BLP} placed
between the ring cyclotron and a scattering chamber. Each polarimeter
utilized left-right asymmetries in p-H scattering from
(CH$_2$)$_n$ foil to determine the vertical transverse component $p_y$
of the beam polarization. The typical beam polarization was 65\%. Then the
beam was transported to a target center in the scattering
chamber. The typical beam spot size on the target during measurements 
was 1~mm in diameter. 
Finally, the beam was stopped by an internal Faraday cup inside the 
scattering chamber in the case of forward-angle measurements. In the
measurements at backward scattering angles, the beam was transported to
another Faraday cup located inside the shielding wall of the
experimental room about 25~m downstream of the scattering
chamber. The integrated beam current was monitored using a current digitizer 
~(Model 1000C) made by BIC~(Brookhaven Instruments Corporation). 
Additionally, the beam current was monitored independently using p-p 
cross sections at the BLPs during the backward-angle measurements. 
Five tin isotope targets~($^{116,118,120,122,124}$Sn) in the form of 
self-supporting metal foils were used for this experiment.
Two different thicknesses were used for each target. 
Thin targets were used for the forward-angle measurements to reduce 
the dead time of the data acquisition system, and thick targets were for the 
backward-angle measurements to increase the yields. The enrichment 
and thicknesses of each target are shown in Table~\ref{tab:table1}.
\begin{table}[h]
\caption{\label{tab:table1} Target enrichment and thicknesses 
of tin isotopes.}
\begin{ruledtabular}
\begin{tabular}{cccc}
Nucleus & Enrichment & Thin & Thick \\
\hline
$^{116}$Sn& 95.5$\%$&10.0mg/cm$^2$& 100.mg/cm$^2$\\ 
$^{118}$Sn& 95.8$\%$&10.0mg/cm$^2$& 100.mg/cm$^2$\\ 
$^{120}$Sn& 98.4$\%$& 5.12mg/cm$^2$&39.9mg/cm$^2$\\ 
$^{122}$Sn& 93.6$\%$&10.5mg/cm$^2$&85.4mg/cm$^2$\\ 
$^{124}$Sn& 95.5$\%$& 5.00mg/cm$^2$&62.7mg/cm$^2$\\ 
\end{tabular}
\end{ruledtabular}
\end{table}

The main contaminants of the targets originated from other tin isotopes. 
The present energy resolution could not separate the elastic scattering 
of other tin isotopes. 
Thus, we analyzed the targets including the contamination from other 
isotopes at all momentum transfer regions. We estimated the error of 
this analysis to be less than 1$\%$ for all cross sections and 
analyzing powers. 
We used an automatic target changer system in this experiment to reduce 
the systematic errors of relative cross sections between isotopes.
We formed a stack of three targets, which were moved vertically every 
2~min to avoid errors due to the drift of the beam direction and 
that of its position on the targets.

We used a high-resolution~(p/$\Delta$p $\sim$37,000) magnetic spectrometer, 
'Grand Raiden'~(GR), which had a Q1-SX-Q2-D1-MX-D2 configuration 
and focal-plane counters for momentum analysis~\cite{GR}. 
The focal-plane counters consisted of two vertical-drift-type 
multi wire drift chambers~(VDCs) and two plastic scintillating counters. 
The momentum of protons scattered from the target was analyzed 
using the GR. The trajectory of the scattered protons from the target 
was reconstructed from the position measurement carried out using 
two sets (X1,U1 and X2,U2) of VDCs, which were placed near the focal plane 
of the spectrometer~\cite{VDC}. Each set had an effective area 
of 120 cm [width]$\times$ 10 cm [height]. 
Two 1~cm-thick plastic scintillating counters (PS1 and PS2) placed behind 
the VDCs were used as triggers and for particle identification.

\begin{figure}
  \includegraphics[height=8.6cm]{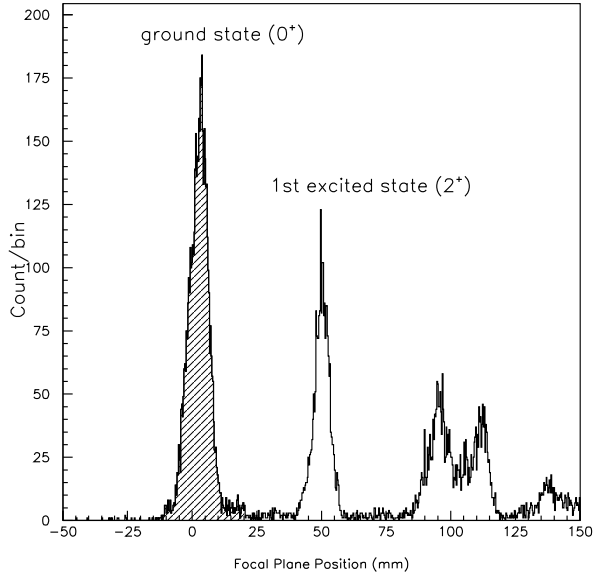}
  \caption{\label{fig:GR1}Sample focal plane spectra corresponding to the 
    excitation energy of $^{120}$Sn, taken at a laboratory scattering 
    angle of 35.5$^\circ$}
\end{figure}

The energy resolution of the beam was 200~keV in FWHM, which was 
determined by the energy width of the beam itself and was sufficient 
to separate elastic peaks from inelastic peaks. Figure~\ref{fig:GR1} 
shows a sample focal-plane spectra of $^{120}$Sn. The differential cross 
sections and analyzing powers were measured for analysis of up to 
50$^\circ$, corresponding to a momentum transfer of 3.5 fm$^{-1}$ 
in the center-of-mass system. We 
included 3\% errors 
for both cross sections and analyzing powers as errors resulting from 
experimental conditions including target thickness uncertainty.

\section{Theoretical Analysis}
\begin{figure}
  \includegraphics[height=8.6cm]{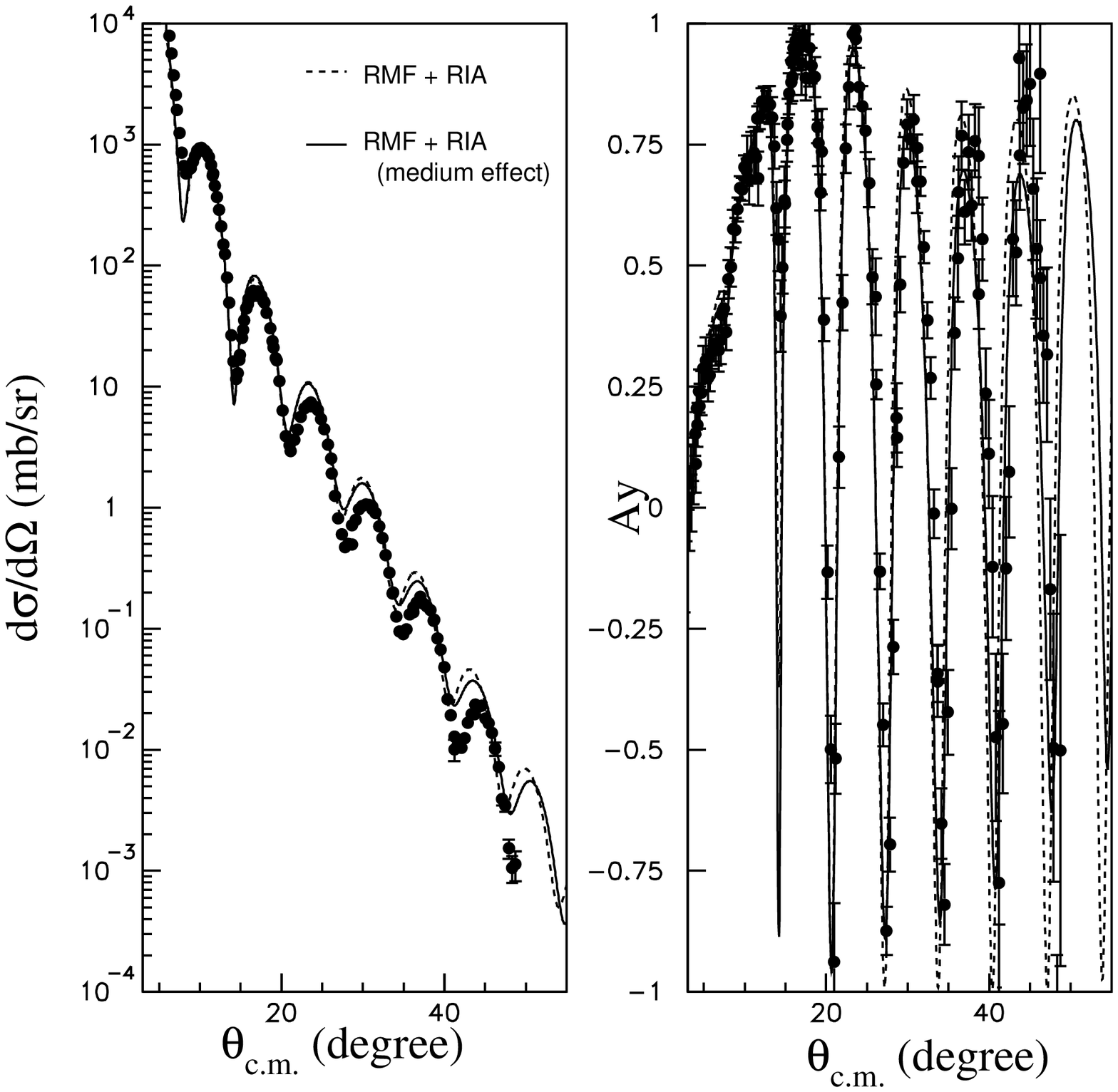}
  \caption{\label{pb}Experimental data for $^{208}$Pb measured at 
    TRIUMF~\cite{TRI} at 300~MeV. Dashed lines show the result of the 
    original RIA calculations using RMF densities~\cite{RH}. 
    Solid lines show the calculation based on
    the modified effective interaction using RMF densities.}
\end{figure}

\
 We analyzed our data using the formula for the relativistic impulse 
approximation~(RIA) using the relativistic Love-Franey 
interaction~\cite{Horo}. The invariant amplitude has been determined 
from nucleon-nucleon phase shifts and is expressed as
\begin{eqnarray}  
F(q) &=& F^S+F^V\gamma^\mu_{(0)}\gamma^\mu_{(1)}
+F^{PS}\gamma^5_{(0)}\gamma^5_{(1)}\nonumber\\
&& +F^T\sigma^{\mu\nu}_{(0)}\sigma_{(1)\mu\nu}
+F^{A}\gamma^5_{(0)}\gamma^\mu_{(0)}\gamma^5_{(1)}\gamma_{(1)\mu}.
\end{eqnarray}
Each amplitude is shown as the sum of real and imaginary amplitudes using 
the masses, coupling constants, and cutoff parameters of exchanged mesons. 
\begin{figure*}
  \includegraphics[height=17.8cm]{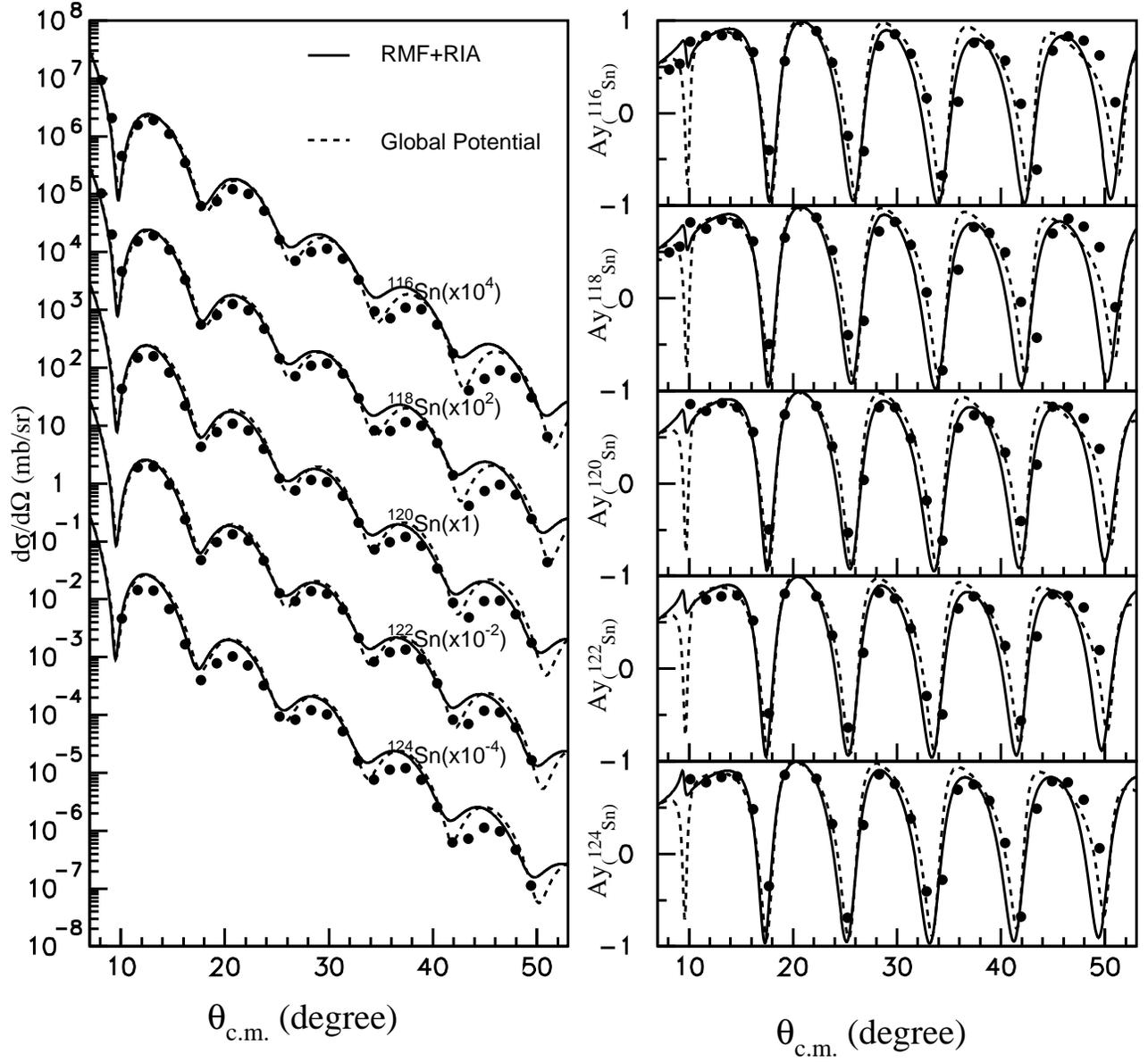}
  \caption{\label{firstcsay}Differential cross sections and analyzing powers 
   for proton elastic scattering from tin isotopes. The solid lines show 
   the results of RIA calculations with the default parameters proposed 
   by Murdock and Horowitz~\cite{Horo} and the RMF densities~\cite{RH}, 
   while dashed lines show calculations 
   using the global potential~\cite{GLO1,GLO2}.}
\end{figure*}
Figure~\ref{firstcsay} shows the experimental results for five tin isotopes 
and two different theoretical results. 
The solid lines in Fig.~\ref{firstcsay} are the results of RIA 
calculations with the default parameters proposed 
by Murdock and Horowitz~\cite{Horo}and the relativistic 
mean field~(RMF) densities~\cite{RH}. 
The dashed lines are the calculations 
using the global potential~\cite{GLO1,GLO2}.
While both calculations are in good agreement with the experimental data 
of obtained from the analyzing powers, the differential cross sections 
are overestimated, particularly in the large-momentum-transfer region. 
Murdock and Horowitz used different masses and coupling constants 
for real and imaginary scattering amplitudes, depending on the interaction. 
In our previous work, we attempted to explain the experimental 
data by phenomenologically changing the masses and coupling constants 
of exchanged mesons~($\sigma,\omega$) in the real and the imaginary 
amplitudes depending on the nuclear density.
The formula for density dependence is as follows;
\begin{eqnarray}
  \label{mf}
 g_j^2, \bar{g}_j^2 &\longrightarrow& \frac{g_j^2}{1+a_j~\rho(r)/\rho_0} 
, \frac{\bar{g}_j^2}{1+\bar{a}_j~\rho(r)/\rho_0} \nonumber \\
 m_j, \bar{m}_j &\longrightarrow& 
 {m_j}[1+b_j~\rho(r)/\rho_0], {\bar{m}_j}[1+\bar{b}_j~\rho(r)/\rho_0],
\end{eqnarray}
where, $m_j, \bar{m}_j, g_j$, {\rm and}, $\bar{g}_j$ indicate the 
masses, and coupling constants of mesons for real and imaginary amplitudes, 
respectively, where {\it j} refers to the $\sigma, \omega$ mesons. 
The normal density $\rho_0$ is 0.1934~fm$^{-3}$~\cite{SAK,Koh}.
These changes in the masses and coupling constants are called 
medium effects, and may be an effect of the presentations of the partial 
restoration of chiral symmetry, Pauli blocking, and multistep processes. 
The tuned effective interaction is applied to existing $^{208}$Pb data 
obtained at TRIUMF~\cite{TRI} with a density distribution calculated from the 
RMF in Fig.~\ref{pb}. 
It was found that the tuning of the effective interaction 
was meaningfully improved compared with the original unmodified interaction. 

For the RIA calculation, we need four density distributions; 
the vector and scalar density distributions of protons and neutrons. 
The proton density distribution and the relation between the 
scalar and vector density can be obtained from the charge distributions, 
the nucleon electric form factors, and the RMF calculation. 
Thus, we can determine the neutron density distribution by comparing 
experimental data with a calculation using the tuned effective interaction. 

In this paper, we attempt to extract neutron density distributions 
for tin isotopes, considering the various ambiguities caused by 
the modification parameters used in the RIA calculation, 
the proton form factors, and an assumption based on scalar densities.

\subsection{Proton density distributions of tin isotopes}
For the RIA calculations we used point proton density distributions 
derived from charge distributions observed in electron-scattering 
experiments~\cite{deVries}. We used the sum-of-Gaussian-type~(SOG)
charge distributions of $^{116,124}$Sn, which were obtained from 
the electron scattering over a 3.5-fm$^{-1}$-wide momentum 
transfer region~\cite{SICK}.
The charge distributions are described as 
\begin{eqnarray}
\rho _{ch}(r) & = & \frac{Z}{2\pi ^{3/2} \gamma ^3 }\sum _{i=1}^{12}
\frac{Q_i }{1+2R_i ^2 /\gamma ^2 }\nonumber\\
&& \times\left[ e^{-(r-R_i)^2/\gamma ^2}+ e^{-(r+R_i)^2/\gamma ^2 } \right].
\label{sogeq}
\end{eqnarray}
However, reported data on charge distributions obtained using 
model-independent distributions are limited. 
\begin{figure}
  \includegraphics[height=8.6cm]{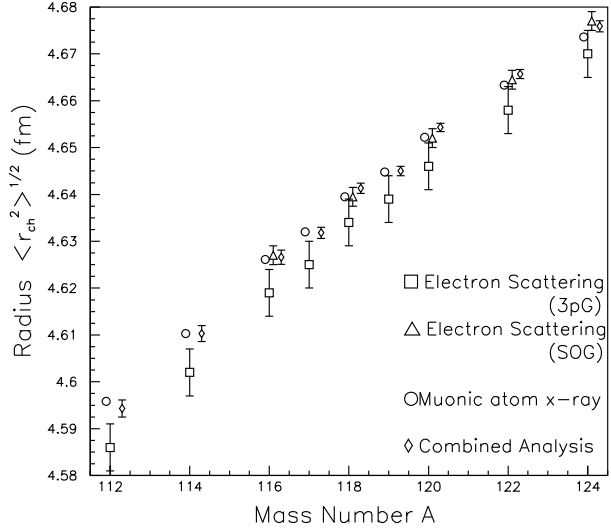}
  \caption{\label{charge}Existing experimental results for charge radius 
    of tin isotopes. Squares and triangles represent results 
    from electron elastic scattering using different shapes of 
    charge distributions~\cite{deVries}. In the case of SOG data, 
    from existing obtained $^{116,124}$Sn data, $^{118,120,122}$Sn data 
    are deduced in this work. 
    Circles show results from muonic x-ray data~\cite{PILL}. 
    Diamonds show combined results from both electron scattering 
    and muonic x-ray~\cite{ANG}}
\end{figure}
Figure~\ref{charge} shows existing experimental results for the 
charge radius of tin isotopes. Since the radius of tin isotopes increases 
smoothly with mass number and the number of protons is constant, 
we can expect a smooth change in the charge distributions 
as the neutron number increases. Thus we obtained the charge distributions 
of $^{118,120,122}$Sn by interpolation using the arithmetic 
mean of the previous derived charge distributions of $^{116,124}$Sn;
\begin{eqnarray}
\rho_{ch}^{^{118}{\it Sn}}(r)&=&\frac{3\rho_{ch}^{^{116}{\it Sn}}(r)+
\rho_{ch}^{^{124}{\it Sn}}(r)}{4}\\
\rho_{ch}^{^{120}{\it Sn}}(r)&=&\frac{\rho_{ch}^{^{116}{\it Sn}}(r)+
\rho_{ch}^{^{124}{\it Sn}}(r)}{2}\\
\rho_{ch}^{^{122}{\it Sn}}(r)&=&\frac{\rho_{ch}^{^{116}{\it Sn}}(r)+
3\rho_{ch}^{^{124}{\it Sn}}(r)}{4}.
\end{eqnarray}
The interpolated radii of model-independent-type charge distributions for 
$^{118,120,122}$Sn are consistent with other experimental data obtained from 
the electron scattering~\cite{deVries} and muonic x-rays~\cite{PILL}, 
as shown in Fig.~\ref{charge}. 
The differences between interpolated radii and other results from 
the combined analyses of elastic electron scattering and muonic x-ray data 
elastic electron scattering and muonic x-ray data are small~\cite{ANG}. 
We estimate that the errors of our deduced charge radii are 
0.003~fm, which are the same order as those for $^{116,124}$Sn

To extract the point proton density distribution we need to unfold 
the charge distribution with the finite size of the protons 
and to consider the contribution from neutrons.
Therefore, we can write a charge distribution $\rho_{ch}$ using the charge 
distributions of a proton $\rho_{ch}^{proton}$ and neutron 
$\rho_{ch}^{neutron}$ as follows;
\begin{eqnarray}
\rho_{ch}(\vec{r})&=&\int \rho_p(\vec{r'})\rho^{proton}_{ch}
(\vec{r}-\vec{r'})d\vec{r'}\nonumber\\
&&+\int \rho_n(\vec{r''})\rho^{neutron}_{ch}(\vec{r}-\vec{r''})d\vec{r''}.
\label{unfold}
\end{eqnarray}
Therefore, we calculate the mean square radius of the charge distribution 
as follows; 
\begin{eqnarray}
\left<r^2_{ch}\right> &=& \left<r^2_{p}\right> 
+ \left<(r_{ch}^{proton})^2\right> 
+ \frac{N}{Z}\left<(r_{ch}^{neutron})^2\right>,
\label{unfoldr}
\end{eqnarray}
where
\begin{eqnarray}
\left<r^2_{ch}\right>&\equiv&\int r^2\rho_{ch}(\vec{r})d\vec{r}
/\int \rho_{ch}(\vec{r})d\vec{r}\nonumber\\
\left<r^2_p\right>&\equiv&\int r^2\rho_p(\vec{r})d\vec{r}
/\int \rho_p(\vec{r})d\vec{r}\nonumber\\
\left<(r^{proton}_{ch})^2\right>&\equiv&\int r^2\rho^{proton}_{ch}(\vec{r})d\vec{r}
/\int \rho^{proton}_{ch}(\vec{r})d\vec{r}\nonumber\\
\left<(r^{neutron}_{ch})^2\right>&\equiv&\int r^2\rho^{neutron}_{ch}(\vec{r})d\vec{r}
/\int \rho^{neutron}_{ch}(\vec{r})d\vec{r}\nonumber.
\end{eqnarray}
$\left<r_{ch}^2\right>^{1/2}, \left<r_p^2\right>^{1/2}, 
\left<(r_{ch}^{proton})^2\right>^{1/2}, {\rm and} 
\left<(r_{ch}^{neutron})^2\right>^{1/2}$ 
are denoted as the root-mean-square (RMS) radii of $\rho_{ch}$, $\rho_p$, 
$\rho_{ch}^{proton}$, and $\rho_{ch}^{neutron}$, respectively.
We used the simple parameterization of nucleon form factors~\cite{KELL}, 
where the RMS radius of the proton itself is 0.863(4)~fm and 
the mean square radius of the neutron itself is -0.112(3)~fm$^2$. 
This proton radius is consistent with that obtained from the measurement 
of the hydrogen 1S Lamb shift~\cite{LAMB}. 
The contribution to the RMS radius of the point proton density 
distribution from the neutron form factor is in the order of 0.02~fm.
Thus, the contribution from the neutron form factor 
is small but not negligible. 

Equation~(\ref{unfold}) shows that we need the neutron density distribution 
to precisely deduce density distribution. 
Since the errors of proton radii are mainly determined by the charge radii 
in the equation, the errors of the RMS radii of the point proton of 
tin isotopes are estimated to be 0.003~fm. 

In this report we tune newly a medium-effect parameter set using the $^{58}$Ni 
data of Refs.~\cite{SAK,TKD}, because we adopted 
a new treatment for the point proton density distribution. 

\subsection{$^{58}$Ni and the effective interaction}
We must calibrate the effective nucleon-nucleon interaction, particularly 
in the nuclear interior, by the scattering from the nucleus. 
In this subsection we discuss the analysis of the elastic scattering 
of $^{58}$Ni at the energy of 295 MeV, which allows to adjust 
the effective interaction further. 

\begin{figure}
  \includegraphics[height=8.6cm]{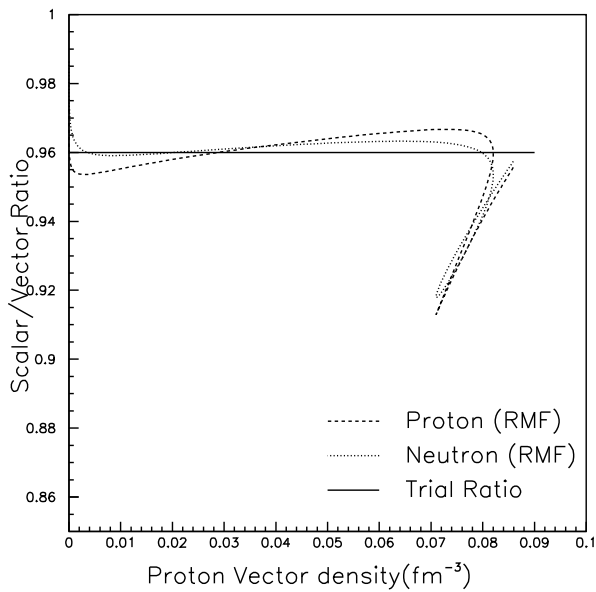}
  \caption{\label{sv2}Ratios of scalar density to vector density 
    for proton and neutron calculated by the RMF calculation for $^{58}$Ni.}
\end{figure}

In the previous work, eight parameters were searched for independently
including the imaginary parts of two exchanged mesons~\cite{SAK}. 
It is difficult to obtain a unique medium-effect parameter set due to 
its many degrees of freedom. This difficulty causes ambiguities in the 
neutron density distributions. Thus, we attempt to express the elastic 
scattering using the medium-effect parameters that have less freedom. 
We need the information of both proton and neutron density distributions to 
calibrate the effective interaction based on RIA calculations. 
Several calculations have been performed to obtain the proton and 
neutron density distributions of $^{58}$Ni using 
relativistic or nonrelativistic mean-field calculations. 
The results are dependent on the interactions used in the calculations 
but the differences between the RMS radii of point protons and neutrons, 
the neutron skin, are generally small~\cite{GOG,LALA,TAG,DOB,HOF,ANT}, 
and are given as
\begin{equation}
\Delta r_{np} \equiv
 \left<r_n^2\right>^{1/2}-\left<r_p^2\right>^{1/2}.
\end{equation}
The used of 800MeV proton elastic scattering based on KMT nonrelativistic 
impulse approximation gave a thickness of 
$\Delta r_{np}$=0.01(5)~fm~\cite{Ray}. 
X-ray measurement of an antiprotonic atom gave a value of 
-0.9(9)~fm~\cite{Trz}.
Thus, in the case of $^{58}$Ni, we can assume that the neutron density 
distribution has the same shape as the proton density distribution. 
Therefore, the neutron density distribution can be described as 
\begin{equation}
\label{ded}
\rho_n(r)=(N/Z)\rho_p(r).
\end{equation}
Thus, the substitution of Eq.~(\ref{ded}) into Eq.~(\ref{unfold}) gives 
Eq.~(\ref{substitute}) the point proton density distribution of $^{58}$Ni, 
\begin{eqnarray}
\label{substitute}
\rho_{ch}(\vec{r})&=&\int \rho_p(\vec{r}')[\rho^{proton}_{ch}
(\vec{r}-\vec{r}')\nonumber\\
  &&+(N/Z)\rho^{neutron}_{ch}(\vec{r}-\vec{r}')]d\vec{r}'.
\end{eqnarray}
According to the RMF calculation, 
the ratio of scalar-to-vector densities has an almost constant value of 
0.96, as shown in Fig.~\ref{sv2}, corresponding to the ratio of the 
integrated scalar density to the integrated vector density. 
Using this constant value, we reproduced the scalar density 
distributions from the RMF calculation.
Even if we had taken different constant value from 0.95 to 0.97, 
the differences in the cross sections and analyzing powers would have 
been less than 2-3\%, which can be compensated for by the ambiguity of 
the medium-effect parameters. 

\begin{figure}
  \includegraphics[height=8.6cm]{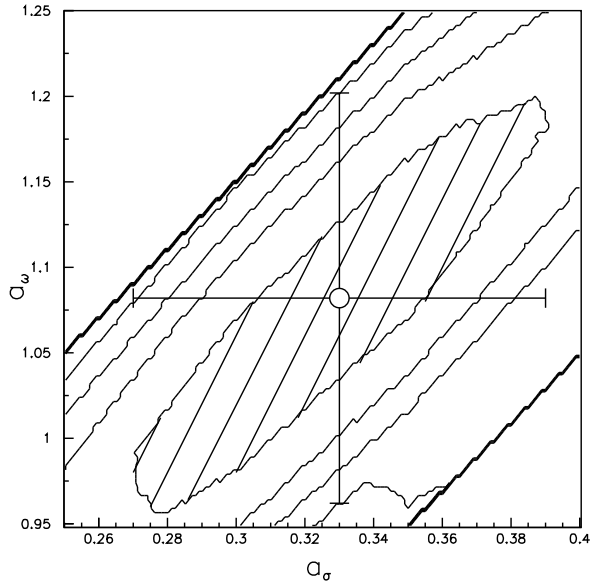}
  \caption{\label{map}Contour plot of $\chi^2$ correlation of 
    a$_\sigma$ with a$_\omega$ for $^{58}$Ni. The hatched area represents 
    the region obtained from Eq.~(\ref{err1}).
    The open circle and bars represent the best-fit parameter and the 
    errors shown in Table~\ref{tab:table2}, respectively. The magnitude 
    of $\chi^2$ is arbitrary.}
\end{figure}

Since in the case of eight parameters, we have too many degrees of freedom 
for fitting parameters and the correlations among these parameters are large, 
in this work we have used the same modification parameters 
for both real and imaginary parts of the scattering amplitude
~($\bar{\rm a}_j$=a$_j$~,$\bar{\rm b}_j$=b$_j$). 
Figure~\ref{map} shows the correlation between modification parameters 
of a$_\sigma$ and a$_\omega$. 
The correlation is valley like and very strong; the correlations 
of the other five combinations of modification parameters also show 
similar behavior. Thus, we determined the modification parameters uniquely.

\begin{figure}
  \includegraphics[height=8.6cm]{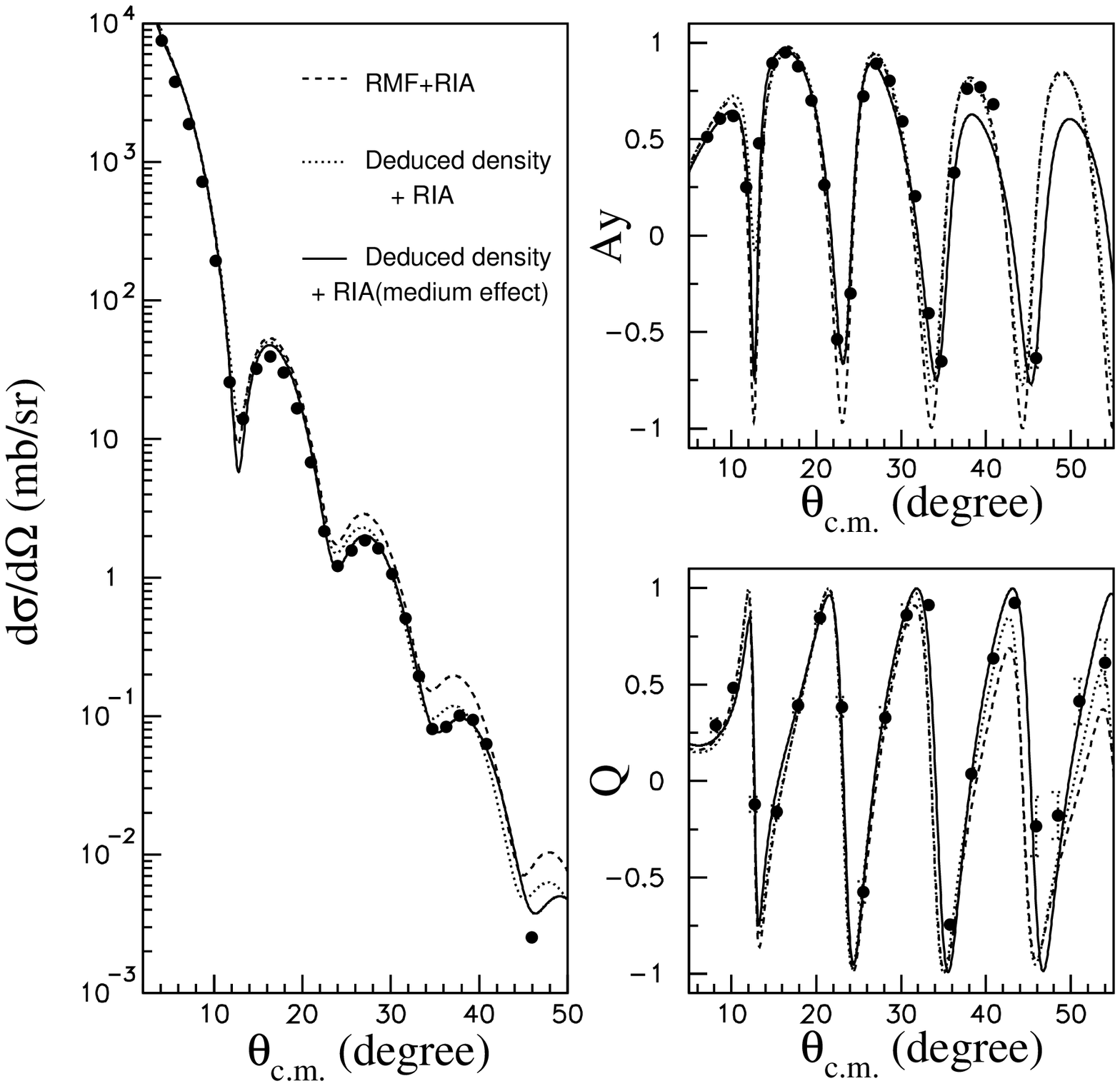}
  \caption{
    \label{niall}
    Experimental data for $^{58}$Ni at 295~MeV and results of several 
    calculations. Dashed lines show results of the original RIA 
    calculations using RMF densities. Dotted lines represent similar 
    calculations but using densities deduced from electron-scattering data. 
    Solid lines show calculations based on the modified 
    effective interaction using the deduced densities.}
\end{figure}

The medium-effect parameters were determined 
by fitting the experimental data so as to minimize the chi-square~($\chi^2$) 
value. $\chi^2$ is defined as 
\begin{equation}
\chi^2 = \sum \left[(x_{exp.}-x_{theo.})/\Delta x_{exp.}\right]^2
\end{equation}
where x$_{exp.}$, $\Delta x_{exp.}$ and x$_{theo.}$ are 
the experimental data, the errors in the data, and the calculation results, 
respectively. Figure~\ref{niall} shows the experimental data at 295~MeV and 
the fits with the RIA calculation using the modified effective 
interaction with the unfolded proton and neutron densities of $^{58}$Ni 
obtained using Eq.~(\ref{ded}). 
Our calculations are in good agreement with all the data of the cross 
sections, the analyzing powers, and the spin rotation parameters for 
up to 3.5 fm$^{-1}$. $\chi^2_{min}$, the minimum of $\chi^2$ for $^{58}$Ni 
data using four free parameters, has almost the same order as that in 
the previous work using eight free parameters.
We estimated the errors of the modification parameters from the statistics 
and the experimental conditions as follows. 
\begin{eqnarray}
\label{err1}
\chi^2 \le \chi^2_{min} + P
\end{eqnarray}
Here, P~(in this work P=4) denotes a number of fitting parameters~\cite{ERR}. 
Table~\ref{tab:table2} shows a summary of the fitting results. 
Figure~\ref{map} shows the correlation between a$_\sigma$ and a$_\omega$. 
The circle shows the best-fit parameter, and horizontal and vertical bars 
show the error bars of a$_\sigma$ and a$_\omega$, respectively.
The hatched area shows the region that satisfies Eq.~(\ref{err1}). 
Since each pair of parameters has a strong correlation, we cannot 
independently use the entire region defined by the errors in 
Table~\ref{tab:table2}, meaning that the usable area defined by each 
pair of the parameter set is very narrow.
 
\begin{table}[h]
\caption{\label{tab:table2}Medium-effect parameters at 295~MeV. 
The error estimation is discussed in Sec.~II.B}
\begin{ruledtabular}
\begin{tabular}{c c c}
$j$ & $\sigma$ & $\omega$ \\ 
\hline
$a_j$ &  0.33 $\pm$ 0.06 &  1.08 $\pm$ 0.12 \\
$b_j$ & -0.12 $\pm$ 0.05 & -0.30 $\pm$ 0.03 \\
\end{tabular}
\end{ruledtabular}
\end{table}

\subsection{Tin isotopes and neutron density distributions}
We have tried to deduce the neutron density distributions of tin isotopes 
using the newly modified effective interaction. The scalar nucleon 
density distributions of tin isotopes were assumed to be 0.96 times 
the vector nucleon density distributions as in the case of $^{58}$Ni, 
because the ratios of the scalar to vector densities for tin isotopes 
are almost the same~\cite{RH}. The initial proton density distributions 
of tin isotopes are obtained by Eq.~(\ref{unfold}) using the neutron density 
distribution given by Eq.~(\ref{ded}).  
We used a SOG for the neutron density distribution, which is expressed as
\begin{eqnarray}
\label{sogn}
\rho _{n}(r) & = & \frac{N}{2\pi ^{3/2} \gamma ^3 }\sum _{i=1}^{12}
\frac{Q_i }{1+2R_i ^2 /\gamma ^2 }\nonumber\\
&& \times\left[ e^{-(r-R_i)^2/\gamma ^2}+ e^{-(r+R_i)^2/\gamma ^2 } \right].
\end{eqnarray}
Here, N is the number of neutrons. 
Equation~(\ref{sogn}) is almost the same as Eq.~(\ref{sogeq}). 
While parameters such as $\gamma$ and $R_i$ were fixed using results of 
the charge distribution of $^{116,124}$Sn~\cite{deVries}, 
Q$_i$ were determined independently under the following normalization 
condition; 
\begin{equation}
\label{norm}
\int \rho _n (r)dr =  N  \Rightarrow \sum _{i=1} ^{12} Q_i =  1.
\end{equation}
For fitting, we used a range of momentum transfer from 0.7 
fm$^{-1}$ to 3.5 fm$^{-1}$, since the data for the forward angle~
($\theta_{\rm c.m.} < 8^\circ$) were difficult to measure experimentally.
Our range covers the same momentum transfer region of electron scattering 
as that in Ref.~\cite{deVries}. For the medium-effect parameters, 
we used the same parameters as those determined by the scattering 
from the $^{58}$Ni target. 

\begin{figure*}
  \includegraphics[height=17.8cm]{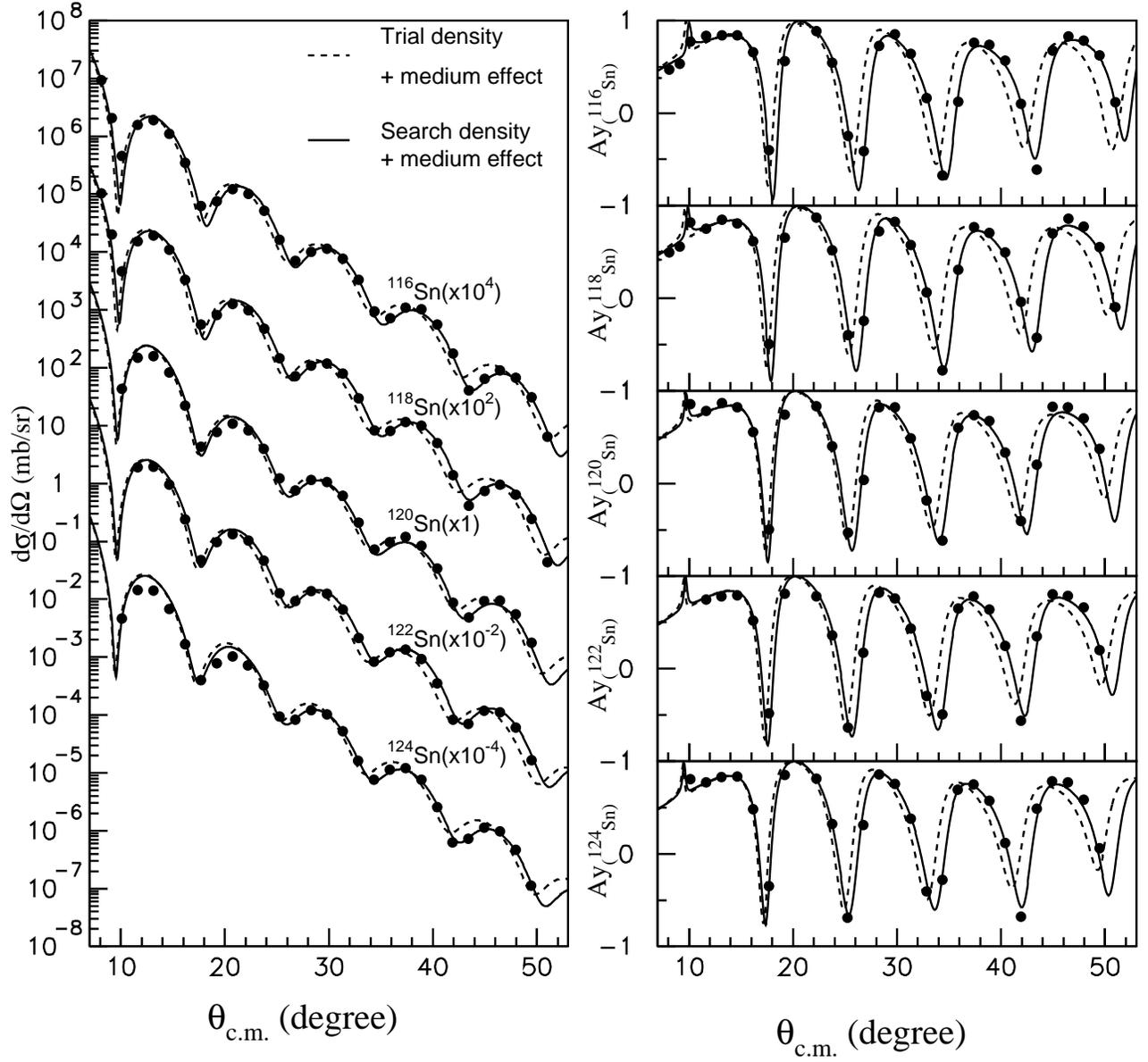}
  \caption{\label{csay}Differential cross sections and analyzing powers 
   for proton elastic scattering from tin isotopes. The figure includes data 
   from Fig.~\ref{firstcsay}. The dashed lines show calculations based on 
   the modified effective interaction using trial density distributions. 
   The solid lines are best-fit calculations based on the modified 
   effective interaction.}
\end{figure*}
 In Fig.~\ref{csay}, we show experimental data together with the 
fitting results and calculations using the initial neutron density
distributions, which are defined as 
\begin{equation}
\rho_n(r)=\rho_p(r) +(\rho_n^{RMF}(r)-\rho_p^{RMF}(r)),
\end{equation}
where $\rho_{n (p)}^{RMF}(r)$ denote the RMF neutron~(proton) density 
distributions~\cite{RH}. We determine the proton and neutron density 
distributions by iterating Eq.~(\ref{unfold}) several times 
until self-consistency is achieved.
The process of iteration hardly affects the proton densities, 
because the proton density distributions are not sensitive to the fine 
structure of the neutron density distributions.

 The calculations using the initial neutron density and the medium-modified 
effective interaction reproduce the absolute value of the cross 
sections and analyzing powers. The angular distributions of both the 
differential cross sections and the analyzing powers using the initial 
neutron density are shifted to forward angle in angle 
because the RMF calculation has a tendency of a larger neutron radius.

\subsection{Uncertainties of neutron density distributions}
We estimated the uncertainties of the neutron density distributions of 
the tin isotopes. There are generally two types of uncertainty. 
One originates from the experiment, and the other originates from 
the model of calculation. In our case, the reduced chi-square 
$\chi^2 / \nu$ is larger than 1, and is typically 7 for 
$^{116,118,122,124}$Sn. In the case of $^{120}$Sn, not only 
the cross section and analyzing powers but also the spin rotation 
parameters~\cite{TKD} are used as experimental data in the analysis. 
Thus, the errors for $^{120}$Sn resulting from the experimental conditions, 
degrees of freedom $\nu$ and $\chi^2 / \nu$ are different from those for 
other isotopes. $\chi^2 / \nu$ is about 10 for $^{120}$Sn. 
If the model were perfect, it would realize $\chi^2 / \nu$=1. 
Thus, the reason that $\chi^2 /\nu > $1 is attributed to the inadequacy of 
our model. To compensate for this inadequacy of the model, 
we increased the errors artificially by multiplying all the experimental 
errors by a constant factor to realize $\chi^2 / \nu$=1. 
This means that we must redefine Eq.~(\ref{err1}) as follows;
\begin{eqnarray}
\label{err2}
\chi^2 \le \chi^2_{min} + P \times {\chi}^2_{min}/\nu. 
\end{eqnarray}
The uncertainties of the neutron density distributions are estimated 
by the following procedure. 

First, we obtain a new medium-effect parameter set using $^{58}$Ni 
data by the Monte Carlo method. We adopt the parameter sets that 
satisfy Eq.~(\ref{err1}) as 'good' medium-effect parameter sets. 

Second, we estimate the uncertainties of neutron density distributions 
using each set of good medium-effect parameters. 
The neutron density distributions are calculated by the Monte 
Carlo method under the normalization condition of Eq.~(\ref{norm}). 
We compare the experimental data with the result 
of the calculation using each good set of medium-effect parameters and 
the trial neutron density distributions. 
We also adopt all neutron density distributions that satisfy 
Eq.~(\ref{err2}) as good neutron density distributions.

We repeated this procedure until the uncertainty of each good neutron 
density distribution converged.
The uncertainties of the neutron density distributions in this analysis are 
defined as the outskirts of the density distributions at each radial point. 
The uncertainties of the RMS radii of the neutron density distributions 
were also calculated using all the neutron density distributions that 
satisfy Eq.~(\ref{err2}). 

\begin{figure}
  \includegraphics[height=8.6cm]{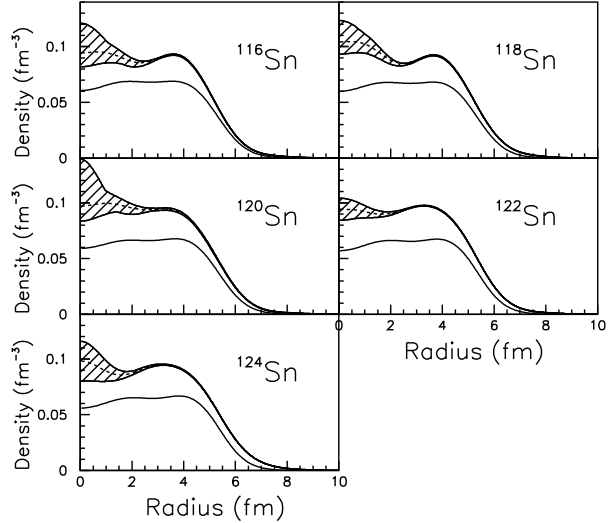}
  \caption{\label{den}Point nucleon density distributions for tin isotopes. 
    Solid lines show point proton density distributions.
    Dashed lines show best-fit neutron density distributions.
  Hatched areas represent the error envelopes encompassing all 
  the trial SOG neutron density distributions deduced by the modified 
  effective interaction. Not only statistical and experimental 
  errors but also systematic errors from the model dependence contribute
  to these regions. }
\end{figure}

Figure~\ref{den} shows the deduced neutron density distributions 
and point proton density distributions of the tin isotopes. Hatched areas 
in the figure represent outskirt regions encompassing all the neutron 
density distributions for tin isotopes allowed by Eq.~(\ref{err2}), 
corresponding to the uncertainties of 
the neutron density distributions in this analysis. 
Table~\ref{tab:table3} shows a summary of the RMS radii of the proton 
and neutron density distributions and the neutron skin thicknesses 
of the tin isotopes. 

\begin{table}
\caption{\label{tab:table3}Reduced RMS radii and thicknesses of tin isotopes.}
\begin{ruledtabular}
\begin{tabular}{c c c c}
Target & $\left<\rho_p^2\right>^{1/2}$~(fm)& $\left<\rho_n^2\right>^{1/2}$ (fm)&~$\Delta r_{np}$~(fm)\\
\hline
$\rm ^{116}Sn$ & 4.562 $\pm$ 0.003&4.672 $\pm$ 0.018&0.110 $\pm$ 0.018\\
$\rm ^{118}Sn$ & 4.575 $\pm$ 0.003&4.720 $\pm$ 0.016&0.145 $\pm$ 0.016\\
$\rm ^{120}Sn$ & 4.589 $\pm$ 0.003&4.736 $\pm$ 0.033&0.147 $\pm$ 0.033\\
$\rm ^{122}Sn$ & 4.602 $\pm$ 0.003&4.748 $\pm$ 0.016&0.146 $\pm$ 0.016\\
$\rm ^{124}Sn$ & 4.615 $\pm$ 0.003&4.800 $\pm$ 0.017&0.185 $\pm$ 0.017\\
\end{tabular}
\end{ruledtabular}
\end{table}

\begin{figure}
  \includegraphics[height=8.6cm]{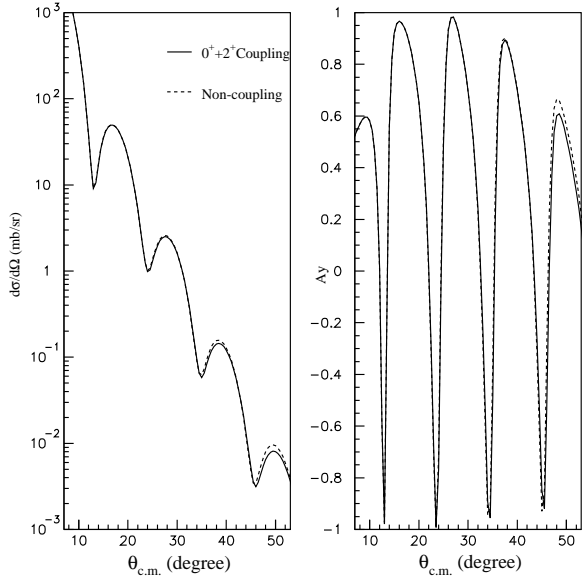}
  \caption{\label{coupling}Coupled-channel effect of $^{58}$Ni at 295~MeV. 
    Solid lines show the coupled-channel calculation between the ground 
    state and the first 2$^+$ state. Dashed lines show the calculation 
    without the coupled-channel effect. The global potential was used 
    as the optical potential. Lower shows that ratios of cross sections 
    and analyzing powers with to without performing the coupled-channel 
    calculation.}
\end{figure}

In the high-momentum-transfer region, the multistep process  
may affect the angular distribution of elastic scattering~\cite{CC}.
To estimate the effect of multistep processes, we performed 
a coupled-channel calculation by using the coupled-channel code ECIS95 
written by Raynal~\cite{ECIS}.
Figure~\ref{coupling} shows the calculated cross sections and 
analyzing powers with and without performing a coupled-channel calculation 
using the global potential~\cite{GLO1,GLO2}. 
The contribution from the coupled-channel calculation for $^{58}$Ni in the 
momentum transfer range from 0.7~fm$^{-1}$ to 3.5~fm$^{-1}$ is 
relatively small (1-2\%), which is less than the uncertainty 
of the medium-effect parameter. In the case of the tin isotopes, 
the situation is similar to the case of $^{58}$Ni.
Thus, the effect from the coupled channel is masked by the uncertainties 
of our introduced medium effects in this analysis. 
Also, the coupled-channel effect might be partly included in our 
parametrization because we aimed to phenomenologically reproduce the 
experimental data. Thus, the effect of the coupled-channel is negligible 
and was not included explicitly in our analysis. 

\section{Results and Discussion}
\begin{figure}
  \includegraphics[height=8.6cm]{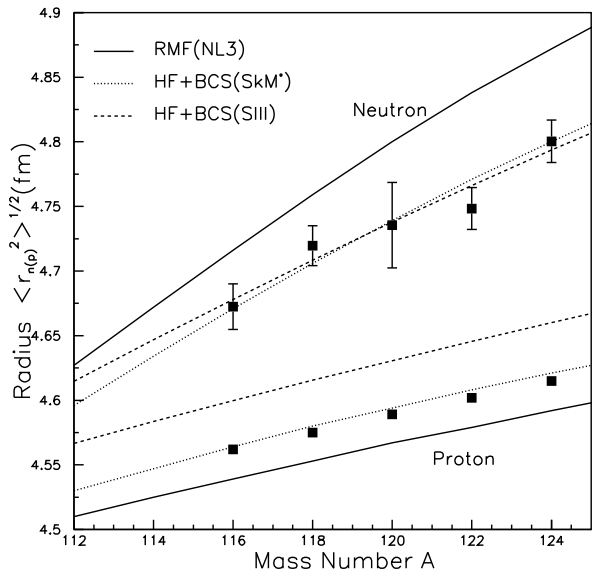}
  \caption{\label{radii}RMS radii of point proton and neutron 
    of tin isotopes. Solid, dotted, and dashed lines show the results of 
    theoretical calculations using typical mean-field models: 
    RMF with NL3~\cite{LALA}, SHF including BCS force 
    with SIII~\cite{TAG}, and 
    with SkM$^*$~\cite{Rein}, respectively.}
\end{figure}

\begin{figure}
  \includegraphics[height=8.6cm]{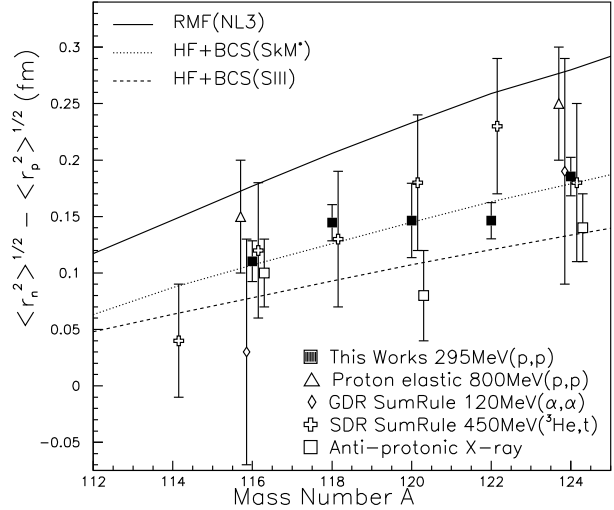}
  \caption{\label{thick}Neutron skin thicknesses of tin isotopes obtained 
    by various methods. Our results are indicated by solid squares. 
    Results from proton elastic scattering at 800~MeV~\cite{Ray}, 
    giant dipole resonance~\cite{Krasz0}, spin dipole 
    resonance~\cite{Krasz}, and antiprotonic x-ray data~\cite{Trz} 
    are shown by open triangles, open diamonds, open crosses, 
    open squares, respectively. The lines represent the models 
    described in Fig.~\ref{radii}.}
\end{figure}

We have obtained the uncertainties of neutron density distributions 
considering statistics, systematic experimental errors, the uncertainties 
of the modification parameters, and our model, as shown in Fig.~\ref{den}. 
Therefore the differences among the deduced shapes of the neutron density 
distributions of tin isotopes gradually changes. 
However the deduced RMS radii of the point proton and neutron density 
distributions clearly increase with the mass number, 
as shown in Fig.~\ref{radii}. 
This tendency is also shown in the results of several typical theoretical 
calculations, which are also plotted in the figure and have slopes 
consistent with our experimental results. 
Among the calculations, the nonrelativistic Skyrme Hartree Fock~(SHF) 
calculations using SkM$^*$ parametrization are in good agreement with 
the RMS radii of both point proton and neutron 
density distributions. Neutron skin thicknesses $\Delta r_{np}$ are 
shown in Fig.~\ref{thick}.
The neutron skin thicknesses are about 0.11-0.19~fm 
for tin isotopes, which are not large what some RMF models suggest. 
The values of our $\Delta r_{np}$ are reproduced by the 
SHF calculation using SkM$^*$ parameterization. 
On the other hand, the values are 
larger than SIII and smaller than NL3. 
Oyamatsu {\it et al.} and Chen {\it et al.} calculated a linear relation 
between the $\Delta r_{np}$ and the symmetry term of the EOS~\cite{OYA,CHE}. 
Therefore, our results favor medium values for the symmetry energy and 
its density dependence used in SkM$^*$ one, 
which are not so large as NL3 but larger than SIII.

\section{Summary}
In this work, we have extracted the neutron density distributions of tin 
isotopes. The experimental data were analyzed in the framework 
of the RIA using a newly tuned relativistic Love-Franey interaction 
obtained for $^{58}$Ni. 
The uncertainties of the neutron density distributions were estimated 
in consideration of not only experimental but also model uncertainties. 
Using the tuned interactions in the nuclear medium, the neutron 
density distributions of the tin isotopes were deduced so as to 
reproduce the experimental data of the isotopes. 
We also deduced the RMS radii of the point proton and neutron density 
distributions. We compared our experimental results with those of several 
theoretical mean-field calculations. SHF calculations using SkM$^*$ 
parameterization as above were in good agreement with the 
RMS radii of both point proton and neutron density distributions. 
We observed a clear increase in neutron skin thickness with mass number, 
although the values obtained were not large what some RMF models suggest. 

\begin{acknowledgements}
We would like to thank Prof.~Hatanaka and the operating crew of RCNP
for their support; for providing a clear, stable, and high-intensity beam
during the experiment; and also for beam-time management.
This experiment was performed at RCNP under program No. E147.
\end{acknowledgements}



\end{document}